\documentclass[
reprint,
superscriptaddress,
nofootinbib,
amsmath,amssymb,
aps,
pra,
twocolumn,
longbibliography,
]{revtex4-1}

\usepackage{hyperref}
\usepackage{natbib}
\usepackage[utf8]{inputenc}    
\usepackage{amsmath,graphicx,amssymb,bbold}
\usepackage{color}
\usepackage{mathtools}
\DeclarePairedDelimiter{\ceil}{\lceil}{\rceil}
\usepackage{graphicx}
\usepackage{dcolumn}

\begin{document}

\newcommand{\hl}[1]{{\color{red}{#1}}}

\def\vec#1{\bm{#1}}
\def\ket#1{|#1\rangle}
\def\bra#1{\langle#1|}
\def\braket#1#2{\langle#1|#2\rangle}
\def\ketbra#1#2{|#1\rangle\langle#2|}

\preprint{APS/123-QED}

\title{Atom-only descriptions of the driven-dissipative Dicke model}

\author{Fran\c{c}ois Damanet}
\affiliation{Department of Physics and SUPA, University of Strathclyde, Glasgow G4 0NG, United Kingdom.}
\author{Andrew J.\ Daley}
\affiliation{Department of Physics and SUPA, University of Strathclyde, Glasgow G4 0NG, United Kingdom.}
\author{Jonathan Keeling}
\affiliation{SUPA, School of Physics and Astronomy, University of St Andrews, St Andrews, KY16 9SS, United Kingdom.}

\date{\today}

\begin{abstract}
We investigate how to describe the dissipative spin dynamics of the driven-dissipative Dicke model, describing $N$ two-level atoms coupled to a cavity mode, after adiabatic elimination of the cavity mode. To this end, we derive a Redfield master equation which goes beyond the standard secular approximation and large detuning limits. We show that the secular (or rotating wave) approximation and the large detuning approximation both lead to inadequate master equations, that fail to predict the Dicke transition or the damping rates of the atomic dynamics.  In contrast, the full Redfield theory correctly predicts the phase transition and the effective atomic damping rates.  Our work provides a reliable framework to study the full quantum dynamics of atoms in a multimode cavity, where a quantum description of the full model becomes intractable.
\end{abstract}

\maketitle

\section{Introduction}
Placing ultracold atoms in a high finesse optical cavity provides an ideal platform to study quantum many body physics out of equilibrium.  As a many body open quantum system, it also provides a severe test for theoretical modelling, as the problem size scales as the square of the Hilbert space dimension, and the Hilbert space dimension grows exponentially with the number of atoms and cavity modes involved.  For this reason, much theoretical work has been restricted to modelling single-mode cavities~\cite{Dimer2007,Baumann2010,Kirton2018:review}, and cases where all atoms behave identically, so that mean-field descriptions can be applied, or permutation symmetry can be exploited.  However, to fully explore many body physics one must move beyond mean-field descriptions, and consider multimode optical cavities~\cite{Gop:Emergent,Gopalakrishnan2010a,Gopal:Frust,Gopalakrishnan:2012cf,Kollar2014,Kollar2016,Torggler2017,Vaidya2018,torggler2018quantum,guo2018sign,guo2018emergent}.  Modelling such systems beyond a semiclassical approximation is a major challenge.  However, a separation of energy scales naturally exists, with fast cavity degrees of freedom coupled to slower atomic motion.  This suggests adiabatic elimination could be used to significantly shrink the Hilbert space.  In this paper we show this is indeed possible, but to capture the resulting dissipative dynamics of atoms requires Redfield theory.

Ultracold atoms in optical cavities provide a versatile platform to study a wide variety of questions about engineering and controlling many-body non-equilibrium systems.  In particular, one can produce controllable coherent atom-cavity interactions by using a Raman driving scheme, where atoms in the cavity scatter light between an external pump laser and the cavity modes~\cite{Dimer2007}.  This controllable interaction can be combined with multimode optical cavities, which support degenerate or near-degenerate families of transverse modes.  This allows tuning the spatial structure of the interactions.  Indeed, experiments have realized this using a tunable-length cavity~\cite{Kollar2014}, which can be used both in the
non-degenerate~\cite{Kollar2016} and nearly degenerate confocal~\cite{Vaidya2018,guo2018sign,guo2018emergent} regime.  This has enabled the creation of tunable-range~\cite{Vaidya2018} sign-changing~\cite{guo2018sign,guo2018emergent} interactions between atoms.   A wide variety of applications of such multimode cavity experiments have been considered.  These include realization of quantum liquid crystalline states~\cite{Gop:Emergent,Gopalakrishnan2010a}, simulating dynamical gauge fields and the Meissner effect~\cite{Ballantine2017:Gauge}, realization of spin-glass phases~\cite{Gopal:Frust}, and creating of associative memories~\cite{Gopalakrishnan:2012cf}.  Related to this last concept there have also been a number of proposals of information processing using such systems, including proposing alternate routes to realize Hopfield associative memory~\cite{Torggler2017}, and to solve specific NP hard problems such as the $N$-queens problem~\cite{torggler2018quantum}.  Other quantum generalizations of the Hopfield associative memory~\cite{Rotondo2018:Hopfield} have also been studied.

Much of the work listed above on multimode cavities has made use of semiclassical equations, describing the amplitude of the cavity modes and the classical spin state of the atoms.  Modelling the full quantum dynamics of the multimode system is a significant challenge, as the multimode structures require keeping track of the quantum state of each atom and each mode of light.  Moreover, since the system is driven and dissipative, a full quantum description generally requires a density matrix approach, or an equivalent stochastic approach.  Since the dynamics of the cavity modes are generally faster than those of the atoms, it would be highly advantageous to eliminate the cavity modes  and consider a master equation for the atomic dynamics only.

To explore the properties of different approximations in deriving an atom-only description, we consider a model for which the correct behavior is well known, namely the open Dicke model~\cite{Dimer2007}.  
This model describes $N$ two-level atoms coupled to a single cavity mode; it has been extensively studied because this model has a ground state transition to a superradiant\footnote{Note the use of the term ``superradiant'' in this paper refers to a ground-state or steady-state of the open system in which there is a macroscopic photon field.  This is distinct from the transient superradiance first discussed by~\citet{dicke54} and reviewed by~\citet{Gross1982a}.  It is also distinct from the steady state superradiant laser~\cite{meiser09,Bohnet2012}. See \citet{Kirton2018:review} for further discussion.} state~\cite{Hepp:Super,Wang:Dicke,hepp73:pra}. While the existence of this ground-state phase transition has historically been questioned~\cite{Rzazewski1975a}, more recent works~\cite{Vukics2012b,Vukics2014,vukics2015fundamental,griesser16,DeBernadis2018:cqed,DeBernadis2018:gauge} suggest such a transitions is indeed possible, but with subtleties regarding gauge choice.  No such issues however occur when considering the driven-dissipative realization of the Dicke model~\cite{Kirton2018:review}, and indeed the phase transition has been seen experimentally~\cite{Baumann2010}.  As a single-mode problem, the behavior of this model is well understood both through mean-field approaches~\cite{Dimer2007,Keeling2010,Bhaseen2012}, as well as through exact approaches based on permutation symmetry~\cite{Xu13,Kirton2017a,shammah2018}.

To derive an atom-only master equation, we consider both the cavity mode and the extra-cavity light as forming a structured bath, with a frequency-dependent density of states.  Despite this structure, it is nonetheless possible to produce a time-local (i.e., Markovian) equation of motion for the system density matrix, as long as the effective damping rate due to coupling to the bath is smaller than the energy scale over which the bath density of states varies.  This holds in the limit of weak enough matter-light coupling, where the Born-Markov approximation holds~\cite{Breuer2002}.  A time-local description means memory effects are neglected, allowing for an efficient computation of the atomic dynamics, while capturing the leading order effects of the cavity loss.   

Directly integrating out the bath, and using the Born-Markov equation leads to an equation known as the Redfield master equation~\cite{Redfield1957a,Bloch1957a}.  Such an equation is not necessarily of Lindblad form~\cite{Lindblad1976b}, and so does not always preserve the positivity of the reduced density matrix for all time~\cite{Davies1974,Duemcke1979a,Whitney2008a}, yielding in some situations negative and/or diverging populations.  The equation is of Lindblad form if the system-bath coupling terms all sample the bath at the same frequency, or if the bath has no structure.  However in most cases (including the problem we consider), this is not true. In order to overcome this potential positivity violation, it is  a common practice to use in addition the secular approximation, introduced by~\citet{Wangsness1953b}. This approximation amounts to neglecting the non-resonant transitions induced by the system-bath dynamics, i.e.\ it removes the coupling between populations and coherences related to states of different energies.

In many cases in  quantum optics, this approximation holds very well --- the energy (or frequency) differences are very large compared with the dynamical frequency scales for evolution of the system, and so the neglected terms have very high frequencies in the interaction picture. Indeed, in the quantum optics literature the rotating wave approximation is used, neglecting all counter-rotating terms in the system-bath coupling, and this has an effect identical to the secular approximation. The master equation can then be put into the standard form~\cite{Gorini1976} which, following Lindblad~\cite{Lindblad1976b}, guarantees the positivity of the density matrix for all times.  However, neglecting the couplings between populations and coherences can have a dramatic effect and completely remove important physical processes. Indeed it is known that, compared to exactly solvable problems, secular master equations can lead to wrong results where nonsecular Redfield theory gives qualitatively correct behavior~\cite{Jeske2014a,Cammack2018,Eastham2016,Dodin2018}. In this paper, we show that the secular approximation is also inadequate to describe the dissipative dynamics of the Dicke model in the thermodynamic limit.

In this paper we present a variety of atom-only descriptions for the open Dicke model, in the form of effective master equations.  These different forms correspond to making or not making the secular approximation, or making an approximation based on the small ratio of atomic energy vs cavity linewidth (i.e., the large bandwidth limit).  We will see that these various approximations significantly modify the attractors of the dynamics, and that only the full Redfield theory correctly captures the known behavior of the driven Dicke model.  Moreover, we will show how semiclassical equations derived from the full model capture dissipative processes which are lost if one first writes semiclassical equations of the Dicke model, and then adiabatically eliminates photons. By comparison to known results we demonstrate that we can derive a master equation for the atom-only system which captures all the required dissipative dynamics.  This provides a firm foundation for future work to model the atom-only dynamics in multimode cavities, making use of advanced numerical methods~\cite{Daley2004:MPDO,Verstraete2004:MPDO,Zwolak2004:MPDO,Finazzi2015:Corner,Jin2016:Cluster}.

The remainder of the article is arranged as follows.  In Sec.~\ref{sec:modelbg} we introduce the open Dicke model, and review the well known behavior of this model, both in terms of its steady states, and the dissipative approach to those states in the limit where atomic energies are much smaller than the cavity linewidth.  Section~\ref{sec:atomonly} then presents the atom-only equations of motion, and discusses the form that these take with and without various approximations.  The results of each of these different approximations are given in Sec.~\ref{sec:dynamics}, giving the exact solution in some cases, and numerical and analytic approximations for the full (unsecularized) model.   Finally, in Sec.\ref{sec:conclusion} we summarize our results, and discuss some potential future applications enabled by this work.

\section{Model and Background}
\label{sec:modelbg}

\subsection{Raman-driven realization of Dicke model}
\label{raman}

\begin{figure}[htpb]
    \centering
    \includegraphics[width=3.2in]{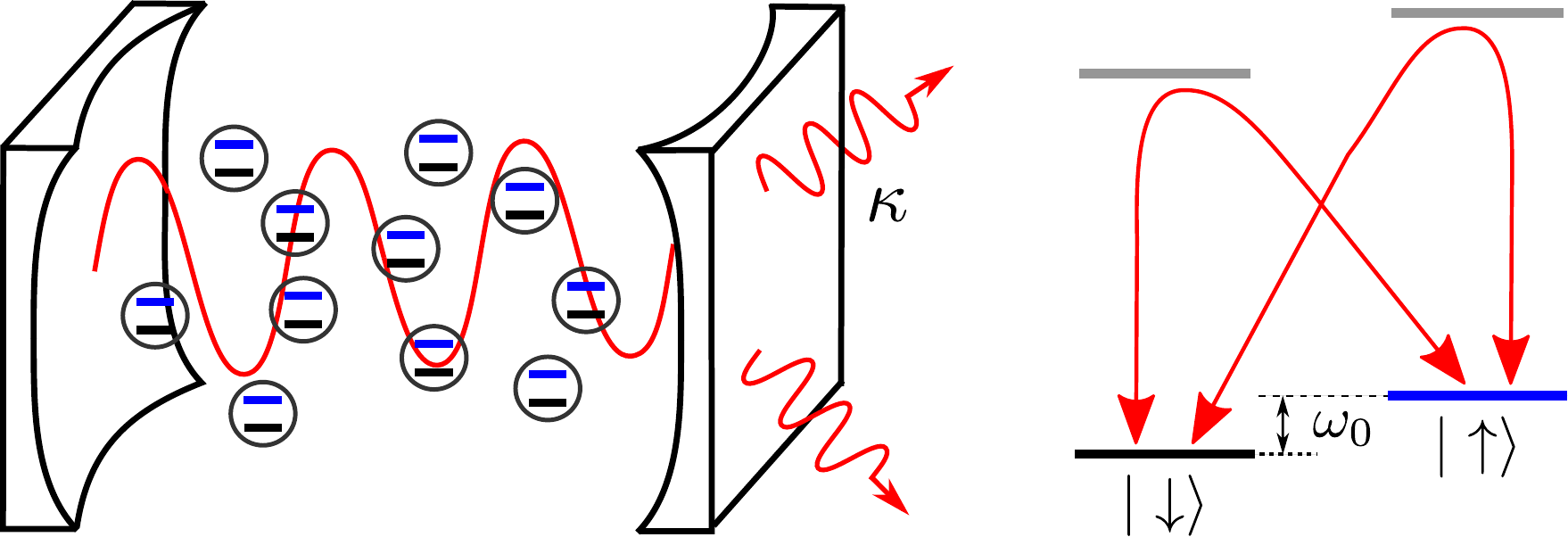}
    \caption{Cartoon of the Dicke model.  Left: Many two-level systems placed in a lossy cavity.  Right: Raman driving scheme; transitions between two states $\ket{\downarrow}$ and $\ket{\uparrow}$ involve a virtual transition due to scattering between a pump and a cavity photon. }
    \label{fig:schematic}
\end{figure}

We consider the Dicke model~\cite{Hepp:Super,Wang:Dicke,hepp73:pra}, describing $N$ identical two-level atoms collectively coupled to a single-mode lossy cavity~\cite{Kirton2018:review}.  As described in Ref.~\cite{Dimer2007}, such a model can be realized as an effective low energy description of atoms with Raman driving.  That is, transitions between two low lying atomic states are driven by scattering a pump photon into a cavity mode, or vice versa (see Fig.~\ref{fig:schematic}).

In this context, working in the rotating frame of the pump, one realizes the Dicke Hamiltonian, combined with optical losses from the cavity mode~\cite{Dimer2007,Bhaseen2012}.   The problem is thus described by the master equation:
\begin{equation}
    \label{eq:orig_master}
    \partial_t \rho 
    =-i[H_{\text{Dicke}},\rho]
    + \kappa \mathcal{L}[ a],
\end{equation}
where $\mathcal{L}[X]=2  X \rho X^\dagger - \{  X^\dagger  X, \rho\}$ describes photon losses with a rate $\kappa$ and where the Dicke Hamiltonian is given by:
\begin{equation}
    \label{eq:DickeH}
    H_{\text{Dicke}}
    = \omega_0 S^z
    + \omega  a^\dagger  a
    + 2 g \left(  a +  a^\dagger\right)
    {S}^x.
\end{equation}
The first term describes the level splitting $\omega_0$ of the two low-lying atomic states, where $S^\alpha = (1/2) \sum_{i = 1}^N \sigma_i^\alpha$ ($\alpha = x,y,z$) are collective spin operators written in terms of the standard single-spin Pauli operators $\sigma_i^\alpha$ ($i = 1,\dotsc, N$). The second term describes the cost of scattering photons into the cavity, where $a$ is the annihilation operator of a cavity photon, and $\omega$  the detuning of the cavity mode from the pump frequency. The final term results from the Raman process, leading to an effective interaction between the atoms and the cavity field, where $g = g_0 \Omega/\Delta_a$  in terms of the bare coupling $g_0$, the Rabi frequency $\Omega$ of the transverse pump and the atomic detuning $\Delta_a$. These definitions are chosen to match the Hamiltonian in Ref.~\cite{Bhaseen2012}, for ease of comparison to the semiclassical results presented there.

\subsection{Review of dynamics of the dissipative Dicke model}
\label{reviewDicke}

For completeness, in this section we briefly summarize the well-known properties of the model described by Eq.~\eqref{eq:orig_master} and Eq.~\eqref{eq:DickeH}.  
In the thermodynamic limit (i.e., large $S$), the behavior of the model is well-described by the semiclassical equations of motion~\cite{Bhaseen2012}:
\begin{equation}
\label{eq:fullsemiclassics}
\begin{aligned}
    \partial_t \langle S^- \rangle &= 
    - i \omega_0 \langle S^- \rangle + 
    2 i g \left(\langle a \rangle  + \langle a^\dagger \rangle\right) \langle S^z \rangle,\\
    \partial_t \langle S^z \rangle &=  2 g 
    \left(\langle a \rangle + \langle a^\dagger \rangle\right) \langle S^y \rangle,\\
    \partial_t \langle a \rangle &= - (\kappa + i \omega) \langle a \rangle - 2 i g \langle S^x \rangle.
\end{aligned}
\end{equation}
Following these equations, one may see this model shows a phase transition between two classes of steady state attractor: normal states, where $\langle S^x \rangle = \langle  a +  a^\dagger\rangle =0$, and an ordered state where these expectations are non-zero.  This ordered state spontaneously breaks the $Z_2$ symmetry of the model under the transformation $S^x\to -S^x,  a \to -  a$.  By analogy to the ground state phase transition in the Dicke model~\cite{Hepp:Super,Wang:Dicke,hepp73:pra}, this ordered state is known as a superradiant state.   For the open system~\cite{Dimer2007,Bhaseen2012} the transition occurs when $g>g_c$ where $4 g_c^2 N = \omega_0 (\omega^2 + \kappa^2)/\omega$.   

The dynamics in both states is dissipative, i.e., there is damped relaxation towards the given steady state. As discussed in detail in~\citet{Bhaseen2012}, this can be characterized by considering the semiclassical equations of motion, and then linearizing around a given steady state.  In general this procedure leads to a quartic equation for the eigenvalues, but this equation can be solved in the limit where $\omega_0 \ll \omega, \kappa$. In the normal state, the eigenvalues of this linearized analysis take the form
\begin{equation}
    \lambda = 
    - \frac{4  \kappa \omega_0 \omega g^2 N}{(\omega^2+\kappa^2)^2}
    \pm i \omega_0 \sqrt{1 - \left(\frac{g}{g_c}\right)^2},
\end{equation}
(see Eq.~(18) of Ref.~\cite{Bhaseen2012}).
Thus, for $g<g_c$, the normal state is absolutely stable, but with a decay rate that can be much smaller than the bare cavity loss rate, particularly in the experimentally relevant regime $\omega_0 \ll \kappa$. When the system becomes superradiant, one must linearize around the new superradiant solution.  This (using Eqs.~(19) and (20) of Ref.~\cite{Bhaseen2012}) gives instead the eigenvalues:
\begin{equation}
%
    \lambda 
    =
    - \frac{ \kappa \omega_0^2}{\omega^2 + \kappa^2}
    \pm i \omega_0 \sqrt{\left(\frac{g}{g_c}\right)^4-1},
\end{equation}
which corresponds to damped oscillations around the steady state.

The results above come from an analysis of the linearized semiclassical equations of the full model i.e., both atomic and photon degrees of freedom, as given in Eq.~\eqref{eq:fullsemiclassics}. If one performs adiabatic elimination of the photon mode on these semiclassical equations, i.e. using 
$\langle a \rangle = - 2 i g \langle S^x \rangle / (\kappa + i \omega)$ one may note that the other equations depend only on the combination 
$\langle a \rangle + \langle a^\dagger \rangle = - 4 g \langle S^x \rangle \omega/(\omega^2 + \kappa^2)$.  Inserting this into the other two equations yields purely conservative dynamics of the collective spin expectation $\langle \mathbf{S} \rangle$.  Indeed, as noted in~\cite{Keeling2010}, this semiclassical spin dynamics corresponds to that from a Lipkin-Meshkov-Glick~\cite{LMG1,LMG2,LMG3} Hamiltonian
\begin{equation}
 H = \omega_0 S^z - \frac{4 g^2 \omega}{\omega^2 + \kappa^2} (S^{x})^2.
\end{equation}
This Hamiltonian does reproduce the existence of a phase transition at the correct $g_c$, but since the dynamics is purely conservative, this atom-only semiclassical theory cannot describe the correct damped decay toward the steady state attractors.  In the following we will see that a correct atom-only semiclassical theory can however be derived by eliminating cavity photons first, and then taking a semiclassical limit.

\section{Atom-only Redfield master equation}
\label{sec:atomonly}

\subsection{Derivation of the Redfield equation}

In this section, we treat the matter-light coupling as a weak system-reservoir coupling and derive a Redfield master equation for the atom-only dynamics.  That is, we derive a description purely in terms of atomic operators, which nonetheless captures the effects of the dissipative cavity mode. 

In order to formally derive the atom-only equations, it is useful to note that the starting model described in Eq.~\eqref{eq:orig_master} can also be written as a purely Hamiltonian problem. That is, one could alternatively describe the same system by enlarging the Hilbert space to describe coupling between the cavity mode and a flat bath of extra-cavity radiation modes, $H_{\text{total}}=H_{\text{Dicke}} +  \sum_k g_k ( a^\dagger  A^{}_k + \text{H.c.}) + \mu_k  A^\dagger_k  A^{}_k$, where the bath spectral density satisfies $\sum_k g_k^2 \pi \delta(\nu_k-\nu)=\kappa(\nu)\equiv \kappa$.  In such a description, adiabatic elimination of the cavity modes means regarding both the cavity and extra-cavity modes as forming the bath.  By diagonalizing these coupled harmonic oscillators, one finds the density of states for this effective bath, and can use this to write the open-system description of the atom-only problem.

Using such an approach, we perform the standard derivation of the Redfield equation, first dividing $H_{\text{Dicke}} = H_0 + H_1$ and then working in the interaction picture with respect to $H_0$. In this interaction picture, the interaction Hamiltonian $H_1$ takes the form:
\begin{equation}
    H_{1}(t)
    = g X(t) [S^+(t) + S^-(t)],
\end{equation}
where  $X(t) = a(t) + a^\dagger(t)$,  and $S^{\pm}(t) = S^{\pm} e^{\pm i\omega_0 t}$ with $S^{\pm} = S^x \pm i S^y$.  The master equation then becomes:
\begin{equation}
    \dot{\rho} = - \int_0^t dt^\prime \text{Tr}_{B}\left(
    \left[ H_{1}(t), \left[ H_{1}(t^\prime), \rho \right] \right]
    \right).
\end{equation}
To evaluate this, we need to find the correlation function of the cavity
photons, thus one needs the time dependence of $a(t)$. This can be found either from the Green's function resulting from diagonalizing the cavity and extra-cavity modes, or alternatively by using Heisenberg-Langevin equations~\cite{Breuer2002} for the cavity modes.  Using the assumed flat spectral density of the extra-cavity modes finds the cavity photon correlation function:
\begin{equation}
    \text{Tr}_B\left( X(\tau) X(0) \rho_{B} \right)
    = e^{-i \omega \tau - \kappa |\tau|}.
\end{equation}
Performing the integrals over time after extending the limit of integration to infinity, we can then write the Redfield equation for the density matrix.  It is convenient to introduce the quantities:
\begin{equation}
    Q_{\pm} = \frac{g^2}{\kappa + i (\omega \pm \omega_0)},
\end{equation}
in terms of which, the Redfield equation in the Schr\"odinger picture takes the form:
\begin{equation} \label{eq:redfield}
\begin{aligned}
\dot{\rho} =& - i \left[ \omega_0 S^z, \rho\right]  - \\  \big\{ &
 Q_{+}  \left( S^+ S^+ \rho - S^+ \rho S^+ \right) + 
 Q_{-} \left( S^+ S^- \rho - S^- \rho S^+ \right) \\ &
 Q_{+} \left( S^- S^+ \rho - S^+ \rho S^- \right) +
 Q_{-} \left( S^- S^- \rho - S^- \rho S^- \right) \\ &
 Q_{-}^{*} \left( \rho S^+ S^+ - S^+ \rho S^+ \right) +
 Q_{-}^{*} \left( \rho S^+ S^- - S^- \rho S^+ \right)  \\ &
 Q_{+}^{*} \left( \rho S^- S^+ - S^+ \rho S^- \right) +
 Q_{+}^{*} \left( \rho S^- S^- - S^- \rho S^- \right)  \big\}.
\end{aligned}
\end{equation}

In writing Eq.~\eqref{eq:redfield}, we have not made the secular approximation. To make this additional approximation we would neglect those terms which are time-dependent in the interaction picture.  In the above equation, it is the terms with two operators $S^-$ or two operators $S^+$ which oscillate at a frequency $\pm 2 \omega_0$ respectively.  
In the following, we will compare a number of different approximations with and without secularization.  To enable this we will introduce a prefactor $\xi$ in front of those terms which are time dependent in the interaction picture, so that $\xi=0$ corresponds to the secular approximation and $\xi=1$ to making no approximation.

\subsection{Master equation in operator form}

The master equation given in Eq.~\eqref{eq:redfield} can be written in a more compact and convenient form:
\begin{align}
    \dot \rho &= -i [H_{\text{eff}}, \rho]
    + \sum_{ij} L_{ij} \left(
    2 C_j \rho C_i^\dagger - 
    \{C^\dagger_i C_j, \rho\}\right),
    \\
    H_{\text{eff}} &= \omega_0 S^z + \sum_{ij} H_{ij}C^\dagger_i C_j.
\end{align}
Here the two-component vector $C$ has components $C_{1,2} = S^\pm$.  To write the Hamiltonian and Lindblad--Kossakowski matrices $H, L$  it is convenient to first define $Q_0$ and $Q_1$ through $Q_\pm = Q_0 \pm Q_1$.   
We then find:
\begin{align}
    H &= 
    Q_0^{\prime\prime} 
    \begin{pmatrix}
    1 & \xi \\ \xi & 1
    \end{pmatrix}
    +
    \begin{pmatrix}
    Q_1^{\prime\prime} &
    i \xi Q_1^\prime \\
    -i \xi Q_1^\prime &
    -Q_1^{\prime\prime}
    \end{pmatrix},
    \\
    L &= 
    Q_0^{\prime} 
    \begin{pmatrix}
    1 & \xi \\ \xi & 1
    \end{pmatrix}
    +
    \begin{pmatrix}
    Q_1^{\prime} &
    i \xi Q_1^{\prime\prime} \\
    -i \xi Q_1^{\prime\prime} &
    -Q_1^{\prime}
    \end{pmatrix},
\end{align}
where $Q_{i}^\prime, Q_i^{\prime\prime}$ refers to real and imaginary parts of these quantities.

The quantities $Q_0$ and $Q_1$ can be thought of as corresponding to the mean and difference of the quantities arising from the co- and counter-rotating terms in the matter-light coupling.  In the following, we will compare this full master equation with the results making a number of commonly used approximations.  Specifically, we will consider the limit $Q_1=0$, which is relevant when $\omega_0 \ll \omega, \kappa$ (i.e. the large detuning limit), and the limit $\xi=0$ corresponding to secularization.  
We next briefly summarize the simplifications that occur to the effective master equation in these various limiting cases:

\subsubsection{Secularized master equation}

In the case $\xi=0$, the Hamiltonian and Lindblad--Kossakowski matrix both become diagonal. As expected from secularization, this latter has positive entries $Q_{\pm}^{\prime} = Q_0^\prime \pm Q_1^\prime = g^2 \kappa/(\kappa^2 + (\omega \pm \omega_0)^2)$ guaranteeing complete positivity. We find the effective Hamiltonian takes the form of a Lipkin--Meshkov--Glick~\cite{LMG1,LMG2,LMG3} model with $XY$ symmetry
\begin{equation}
\label{eq:HLMGsecular}
H_{\text{eff}} = (\omega_0 -  2 Q_1^{\prime\prime}) S^z 
+ 2 Q_0^{\prime\prime} \left[(S^x)^2 + (S^y)^2\right],
\end{equation}
accompanied by simple spin raising and lowering rates:
\begin{equation}
\label{eq:mastersecular}
\partial_t \rho = - i [H_{\text{eff}},\rho]
+ Q_+^\prime \mathcal{L}[S^+]
+ Q_-^\prime \mathcal{L}[S^-].
\end{equation}
Because the effective Hamiltonian has $XY$ symmetry, it conserves the number of excited spins.  This conservation is an expected consequence of secularization, as the interaction picture with respect to $H_0=\omega_0 S^z$ will give time dependence to any term that is not diagonal in the $S^z$ basis. As we will show below, this means the steady state density matrix is diagonal in the $S^z$ basis, and we find particularly simple steady states arising from the competition of the spin raising and lowering processes.

If we combine this secular limit with the large detuning limit where $Q_1$ may be neglected, the equation simplifies further, giving equal rates $Q_\pm^\prime = Q_0^\prime$ for spin raising and lowering processes.

\subsubsection{Large detuning limit}

If we consider the limit where $Q_1$ may be neglected, but avoiding secularization (so $\xi=1$), we also find a simple form of the master equation. In this case:
\begin{align}
    H_{\text{eff}}&=\omega_0 S^z + 4 Q_0^{\prime\prime} (S^{x})^2, \\
    \label{eq:LargeOmegaMaster}
    \partial_t \rho &= 
    i [H_{\text{eff}},\rho] + 4 Q_0^\prime \mathcal{L}[S^x].
\end{align}
In this case, despite the lack of secularization, we still find a completely positive master equation~\cite{Duemcke1979a}.  This is not surprising, as dropping $Q_1$ corresponds to neglecting the energy differences $\pm\omega_0$, so that all operators sample the bath at the same frequency.  As also discussed further below, this equation also has a simple steady state --- since the jump operator is Hermitian, the steady state is a fully mixed density matrix.

\subsubsection{Full model}

While the two limiting cases mentioned above lead to completely positive master equations, this is not true for the full model. We may see this directly by considering the eigenvalues of the Lindblad--Kossakowski matrix $L$.
Specifically,  the eigenvalues of this matrix are $Q_0^\prime \pm \sqrt{ (Q_0^\prime)^2 + |Q_1|^2}$ which indeed involves a negative eigenvalue for any non-zero $Q_1$.  However, as we will show below, despite this non-positivity, this full master equation is capable of describing the known behavior of the open Dicke model.  This is in contrast to both the limiting cases which cannot reproduce the known behavior at the superradiance transition.

\subsection{Master equation in the Dicke basis}

Since the master equation is written only in terms of collective spin operators, it is convenient to write the master equation in the Dicke basis spanned by the Dicke states $\ket{SM}$ with $S = N/2$ and $M = -S,\dotsc, S$ which satisfy:
\begin{equation}
    S^z  \ket{SM} = M \ket{SM} \qquad
    S^\pm \ket{SM} = f_\pm^M \ket{SM\pm 1},
\end{equation}
where $f_\pm^{M} = \sqrt{(S\mp M)(S \pm M +1)}.$ In this Dicke basis, the density matrix can be decomposed as
\begin{equation}
    \rho(t) = \sum_{M = -S}^{S} \rho_{M,M^{\prime}}(t) \ket{SM}\bra{SM^{\prime}},
\end{equation}
with $(N+1)^2$ matrix elements given by $\rho_{M,M^{\prime}}(t) = \bra{SM} \rho(t) \ket{SM^{\prime}}$.
Noting that $f_-^M = f_+^{M-1}$ we use only $f_+^M$ and suppress the subscript $+$ from hereon.  
The master equation~(\ref{eq:redfield}) for these matrix elements reads
\begin{equation}\label{eq:DickeBasisMaster}
    \begin{aligned}
    &\dot{\rho}_{M,M^{\prime}} = -i \omega_0 (M-M^{\prime}) \rho_{M,M^{\prime}}  - \\
     \bigl\{&  \xi Q_+ \left( f^{M-1}f^{M-2} \rho_{M-2,M^{\prime}} - f^{M-1} f^{M^{\prime}} \rho_{M-1,M^{\prime}+1} \right) + \\
    & Q_- \left( f^{M-1}f^{M-1} \rho_{M,M^{\prime}} - f^{M} f^{M^{\prime}} \rho_{M+1,M^{\prime}+1}\right) + \\
    & Q_+ \left( f^{M}f^{M} \rho_{M,M^{\prime}}  - f^{M-1} f^{M^{\prime}-1} \rho_{M-1,M^{\prime}-1}\right) + \\
    & \xi Q_- \left( f^{M}f^{M+1} \rho_{M+2,M^{\prime}}  - f^{M} f^{M^{\prime}-1} \rho_{M+1,M^{\prime}-1} \right) + \\
    & \xi Q_-^* \left( f^{M^{\prime}+1}f^{M^{\prime}} \rho_{M,M^{\prime}+2} - f^{M-1} f^{M^{\prime}} \rho_{M-1,M^{\prime}+1}\right) + \\
    & Q_-^* \left( f^{M^{\prime}-1}f^{M^{\prime}-1} \rho_{M,M^{\prime}}  - f^{M} f^{M^{\prime}} \rho_{M+1,M^{\prime}+1} \right) + \\
    & Q_+^* \left( f^{M^{\prime}}f^{M^{\prime}} \rho_{M,M^{\prime}}  - f^{M-1} f^{M^{\prime}-1} \rho_{M-1,M^{\prime}-1}\right) + \\
    & \xi Q_+^*  \left( f^{M^{\prime}-2}f^{M^{\prime}-1} \rho_{M,M^{\prime}-2} - f^{M} f^{M^{\prime}-1} \rho_{M+1,M^{\prime}-1} \right)
    \bigr\}.
    \end{aligned}
\end{equation}

\subsection{Atom-only semiclassical dynamics}

In the following sections, to understand the behavior in the thermodynamic limit, it is useful to write semiclassical equations of motion, found by writing equations for
$\langle S^\alpha \rangle$ and then replacing $\langle S^\alpha S^\beta \rangle \to \langle S^\alpha \rangle \langle S^\beta \rangle$.
It is generally easier to extract the equation of motion directly from Eq.~\eqref{eq:redfield}, but introducing the factors of $\xi$. Considering a general operator $A$ we find:
\begin{multline}\label{eq:A}
    \partial_t \langle A \rangle
    = - i \omega_0 \langle [A,S^z] \rangle
    - \Big<
    Q_+ [A,\xi S^+ + S^-] S^+ 
    \\+
    Q_- [A, S^+ + \xi S^-] S^-
    +
    Q_-^\ast S^+ [\xi S^+ + S^-,A]
    \\+
    Q_+^\ast S^- [S^+ + \xi S^-,A]
    \Big>.
\end{multline}
We need only consider equations for $\langle S^z\rangle$ and $\langle S^-\rangle$, since $\langle S^+\rangle$ follows by complex conjugation. We thus find:
\begin{multline}\label{eq:sz}
    \partial_t \langle S^z \rangle
    =    - \Big<
    Q_+ (\xi S^+ - S^-) S^+ 
    +
    Q_- (S^+ - \xi S^-) S^-
    \\+
    Q_-^\ast S^+ (-\xi S^+ + S^-)
    +
    Q_+^\ast S^- (-S^+ + \xi S^-)
    \Big>.
\end{multline}
and
\begin{multline}\label{eq:sm}
    \partial_t \langle S^- \rangle
    = - i \omega_0 \langle S^- \rangle
    + 2 \Big<
     Q_+ \xi S^z S^+ 
    + Q_- S^z S^-
    \\-
    Q_-^\ast S^+ \xi S^z  
    -
    Q_+^\ast S^- S^z
    \Big>.
\end{multline}
Further simplification of these equations depends on whether $\xi=0$ or $\xi=1$.

\section{Spin dynamics of atom-only model}
\label{sec:dynamics}

Having introduced the general model in the previous section, in this section, we analyse the dissipative spin dynamics of this model and each of its limiting cases. In several of these cases, we can exactly solve the model in closed form.

\subsection{Secularized master equation}

For the secularized case, $\xi = 0$, the populations $\rho_{M,M}(t)$ are decoupled from the coherences $\rho_{M,M^\prime \neq M}(t)$, as can be seen from Eq.(\ref{eq:DickeBasisMaster}). Physically, this corresponds to the fact that the Hamiltonian is diagonal in the $S^z$ basis, and the dissipative terms only create or destroy excitations. The equations for the populations (i.e. diagonal elements, $P_M = \rho_{M,M}$) read:
\begin{equation}
 \begin{aligned}
    \dot{P}_M = 
      &2 Q_-^\prime \Big[ (S+M+1)(S-M) P_{M+1} \Big. \\
      \Big.   &\qquad\qquad - (S+M) (S-M+1) P_M
       \Big] \\
       &+
        2 Q_+^\prime  \Big[
        (S+M)(S-M+1) P_{M-1} \Big.  \\
        \Big. & \qquad\qquad - (S+M+1) (S-M) P_M
        \Big].
    \end{aligned}
\end{equation}
One may solve this explicitly for the steady state using a detailed balance condition, where gain and loss terms must be equilibrated~\cite{scully97}. The only consistent way of doing this consists in equating the 1st and 4th terms. This implies the 2nd and 3rd must then also match, and moreover, one may see that the 2nd and 3rd terms relate to the 1st and 4th by replacing $M \to M-1$. Thus, the only required condition is $P_{M+1}/P_M = Q_+^\prime/Q_-^\prime$. This condition is identical to that for a thermal equilibrium magnet of moment $\mu$ in a Zeeman magnetic field $B$, for which $P_M \propto \exp(M \beta \mu B)$, but with the replacement $\beta \mu B \to \ln(Q_+^\prime/Q_-^\prime)$. We can thus use the standard results of such a model and obtain $\langle S^z \rangle = S B_S(x)$,
where $x = S \ln \left( Q_+^\prime/Q_-^\prime\right)$ and $B_S(x)$ is the Brillouin function~\cite{Fazekas1999}:
\begin{equation}\label{eq:Szsecu}
    B_S(x) = \frac{2S+1}{2S}\coth\left( \frac{2S+1}{2S} x \right)
        - \frac{1}{2S}\coth\left( \frac{1}{2S} x \right).
\end{equation}
In the limit $S \to \infty$ the existence of the factor of $S$ in the argument of the Brillouin function means there is a sharp dependence of the result on the ratio $Q_+^\prime/Q_-^\prime$.  Namely, $\langle S^z \rangle = \pm S$ depending on whether $Q_+^\prime > Q_-^\prime$ or vice versa.  Intermediate values of $S$ only occur when  $|Q_+^\prime/Q_-^\prime - 1| \lesssim 1/S$, a vanishing region at large $S$.

We thus see that in this approach, we never describe a superradiant state, but instead have a state purely determined by the ratio of spin flip rates.   This is consistent with our observation from the effective Hamiltonian: the Hamiltonian conserved number of excitations, so could not modify the effects of gain or loss. It is notable that if one considered the effective Hamiltonian on its own, the ground state of this Lipkin-Meshkov-Glick model does have a ground-state phase transition when $Q_0^{\prime\prime}$ is negative.  That is, for $2 |Q_0^{\prime\prime}| N > \omega_0 - 2 Q_1^\prime$,  there is a transition to a state with a finite component of spin in the $xy$ plane.  The effects of dissipation however destroy this transition, and leave only a transition between states aligned along $+z$ and $-z$ axes.

One may also verify that in this limit, the semiclassical equations~(\ref{eq:sz}) and (\ref{eq:sm}) support this result.  The equations for $\langle S^- \rangle$ becomes
\begin{multline}
    \partial_t \langle S^- \rangle
    = \- i \omega_0 \langle S^- \rangle
    + 2 \Big<
    Q_- S^z S^-
    -
    Q_+^\ast S^- S^z
    \Big>.
\end{multline}
which we may rewrite by symmetrizing expressions in the second term as:
\begin{equation}
    \partial_t \langle S^- \rangle
    = - (i \omega_0  + Q_--Q_+^\ast)\langle S^- \rangle
    +  
    2(Q_--Q_+^\ast) \langle S^z\rangle \langle S^-\rangle.
\end{equation}
This can be seen to describe overdamped oscillations of $\langle S^- \rangle$. Regardless of $\langle S^z\rangle$, we always find the steady state obeys $\langle S^- \rangle =0$.
The equation for $\langle S^z\rangle$ similarly becomes
\begin{equation}
    \partial_t \langle S^z \rangle
    = 
    2(Q^\prime_+ - Q^\prime_-) \left(S^2 - \langle (S^z)^2 \rangle \right)
    -
    2(Q^\prime_+ + Q^\prime_-) \langle S^z \rangle.
\end{equation}
In the large $S$ limit, the second term can be neglected, and we find the only steady state is $\langle S^z \rangle = \pm S$, corresponding to the sharp switch noted above.  

If we combine the secular limit with the large detuning limit, where $Q_\pm = Q_0$, we immediately find all probabilities must be equal and so normalization implies $P_M = 1/(N+1)$, i.e. a fully mixed state.  Thus, in this case $\langle S^z\rangle = 0$ independent of all parameters.

\subsection{Large detuning unsecularized master equation}

In the large detuning limit $Q_\pm = Q_0, Q_1=0$ (but without secularization), the decoupling of populations and coherences no longer occurs, i.e. the terms in Eq.~\eqref{eq:DickeBasisMaster} with $M^\prime=M$ couple to other terms with $M^\prime \neq M$.  However, the form of the Master equation in Eq.~\eqref{eq:LargeOmegaMaster} suggests the solution nonetheless remains straightforward.  Namely, since the jump operator in the master equation is Hermitian, a general result~\cite{Breuer2002} states that for Hermitian jump operators, $\rho\propto\mathbb{1}$ is a steady state\footnote{The proof follows from the fact that the identity always commutes with the Hamiltonian, $[ H, \mathbb{1}]=0$, and the identity can make the Lindblad form vanish, i.e. $2 X \mathbb{1}  X^\dagger - \{ X^\dagger  X, \mathbb{1}\}=0$ if ${X}^\dagger={X}$.}.  In such a state we find all expectations of $\langle S^\alpha \rangle$ vanish, and there is no transition as a function of parameters, as obtained in~\cite{Imai2019}.

\subsection{Full model}

If we consider the full model with $\xi=1$ and $Q_1 \neq 0$,
no simple solution exists. Nevertheless, the total number of connected density matrix elements containing the populations is $\ceil[\big]{(n+1)^2/2}$ (since Eq.~(\ref{eq:DickeBasisMaster}) connects matrix elements according to a chequerboard pattern), i.e.\ grows only quadratically with the number of spins, which makes it possible to solve the equations numerically for moderate $N$. In the large $N$ limit, we may also use the semiclassical equations of motion to obtain an analytic expression of the expectation value of the collective spin. In this thermodynamic limit, we find, in contrast to the previous two cases, that there is a superradiant transition. 

In the case $\xi=1$ the semiclassical equations simplify considerably, since we may in general write:
\begin{multline}
    \partial_t \langle A \rangle
    = - i \omega_0 \langle [A,S^z] \rangle
    - 2\left< [A,S^x] (Q_+ S^+  + Q_- S^-) \right>\\
    - 2\left< (Q_+ S^+  + Q_- S^-)^\dagger [S^x,A] \right>.
\end{multline}
In this case it is clearest to write equations for $\langle S^{\alpha=x,y,z}\rangle$ explicitly, rather than for $\langle S^\pm \rangle$.  If we symmetrize all products of operators before taking expectations (i.e. writing $\langle  A  B \rangle = \frac{1}{2} \langle \{ A,  B\} + [ A,  B]\rangle =
\langle  A \rangle \langle  B \rangle + \frac{1}{2} \langle [ A,  B] \rangle$), we then find the following equations:
\begin{align}
    \partial_t \langle S^x \rangle
    &=
    -\omega_0 \langle S^y \rangle,
    \\
    \partial_t \langle S^y \rangle
    &=
    \omega_0 \langle S^x \rangle
    - 4 Q_0^{\prime} \langle S^y \rangle 
    - 8 Q_0^{\prime\prime} \langle S^z\rangle\langle S^x\rangle
    \nonumber\\
    &\qquad\qquad\ - 4 Q_1^{\prime\prime} \langle S^x \rangle
    -
    8  Q_1^{\prime} \langle S^z\rangle\langle S^y\rangle,
    \\
    \partial_t \langle S^z \rangle
    &=
    - 4 Q_0^{\prime} \langle S^z \rangle 
    + 8 Q_0^{\prime\prime} \langle S^x\rangle\langle S^y\rangle
    + 8 Q_1^{\prime} \langle S^y \rangle^2.
\end{align}
We may note that not all these terms are extensive in the thermodynamic limit, where we assume $g^2 N$ is finite so $g^2 \propto 1/N$. Specifically, those terms involving $Q_i$ multiplying a single spin operator (the terms arising from commutators) scale as $g^2$ and so vanish in the limit $N \to \infty$.  In contrast, those terms multiplying two spin operators scale as $g^2 N$ and so remain finite in this limit.   Neglecting such terms is therefore consistent with considering the semiclassical (i.e. mean-field) limit, where fluctuations are suppressed by $1/N$. Since real experiments involve finite numbers of atoms and finite cavity volumes, the practical distinction is that some of the terms in this equation are $N$-fold smaller, and for a typical $N \simeq 10^5$, that difference is significant.  Neglecting these smaller terms then gives a simplified equation of motion:
\begin{equation}\label{eq:semiclassicalfinal}
    \begin{aligned}
    \partial_t \langle S^x \rangle
    &= - \omega_0 \langle S^y \rangle
    \\
    \partial_t \langle S^y \rangle
    &=
   \omega_0 \langle S^x \rangle
    - 8 Q_0^{\prime\prime} \langle S^z \rangle \langle S^x \rangle
    - 8 Q_1^{\prime} \langle S^z \rangle \langle S^y \rangle
    \\
    \partial_t \langle S^z \rangle
    &=
     8 Q_0^{\prime\prime} \langle S^x \rangle \langle S^y \rangle
    + 8 Q_1^{\prime} \langle S^y \rangle^2.
    \end{aligned}
\end{equation}
In the limit where $\omega_0 \ll \omega, \kappa$, but where it remains finite, we may approximate that the two combinations of $Q_i$ appearing here are:
\begin{equation}
    Q_0^{\prime\prime} \simeq
    - \frac{g^2 \omega}{\omega^2 + \kappa^2}, \quad
    Q_1^\prime \simeq
    - \frac{ 2 g^2 \kappa \omega \omega_0}{(\omega^2 + \kappa^2)^2}.
\end{equation}

It is notable that this procedure (eliminating photon modes from the quantum theory, and then deriving the semiclassical equations) does not match the result derived in~\cite{Keeling2010} and reviewed at the end of Sec.~\ref{reviewDicke}.  Namely, if one first makes a semiclassical approximation for the full model, and then eliminates the photon mode, the resulting equations do not match Eq.~\eqref{eq:semiclassicalfinal}; such equations are missing the term proportional to $Q_1^\prime$ which describes damping.  Thus, the approach described here of eliminating photons first and making a semiclassical approximation second appears to restore the missing damping. We discuss the consequences of this in the remainder of this section.

\subsubsection{Steady state.}

To first check the semiclassical theory, we consider the steady state and its comparison to exact solution of the full atom-only model.
The steady state of Eq.~\eqref{eq:semiclassicalfinal} is satisfied by $\langle S^y\rangle=0$ along with $\langle S^x\rangle (\omega_0 - 8 Q_0^{\prime\prime} \langle S^z \rangle)=0$.  This indeed describes two distinct states, a normal state with $\langle S^x \rangle=0$, or a superradiant state that becomes possible once $4 N |Q_0^{\prime\prime}| > \omega_0$, allowing a solution with
$\langle S^z \rangle > -N/2$.   Using the above result for $Q_0^{\prime\prime}$ we indeed see this expression matches the location of the superradiance transition in the full Dicke model.
Using this definition of threshold $4 g_c^2 N = \omega_0 (\omega^2 +\kappa^2)/\omega$  we see that above threshold the solution is
\begin{equation}\label{ssvalues}
   \langle S^x \rangle = \pm \frac{N}{2}\sqrt{1 - \frac{g_c^4}{g^4}}, 
   \qquad
   \langle S^z \rangle = -\frac{N}{2} \frac{g_c^2}{g^2}.
\end{equation}

\begin{figure}
    \centering
    \includegraphics[width=3.2in]{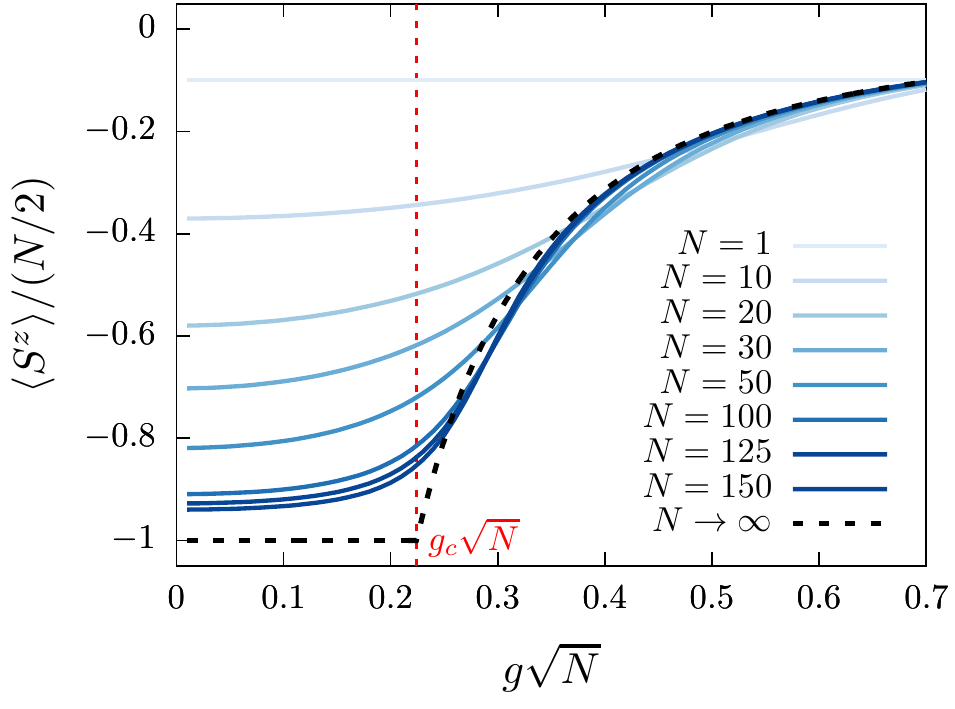}
    \caption{Evolution of the steady state value of $\langle S^z \rangle$ with matter-light coupling $g\sqrt{N}$ for $\omega_0 = 0.1$, and $\kappa = 1$ (in units choosen so that $\omega=1$). Curves are rescaled by $N/2$ so as to produce a finite limit as $N \to \infty$. The blue curves show the exact numerical simulations [based on the resolution of the master equation Eq.~(\ref{eq:DickeBasisMaster})] for various values of $N$ as indicated in the legend.
    The dashed black curve shows the semiclassical solution (valid as $N \to \infty$). The position of $g_c \sqrt{N}$, which marks the transition between the normal state and the superradiant state, is indicated by a red vertical dashed line.}
    \label{fig:comparescexact}
\end{figure}

Figure~\ref{fig:comparescexact} shows the semiclassical solution for $\langle S^z \rangle$ in comparison to results of the exact numerics for finite size calculations. We see that these match well in the large $N$ limit, and that in that limit, a sharp cusp develops in the exact solution.  We focus on $\langle S^z\rangle$, as the finite size calculation never shows symmetry breaking, however, as discussed elsewhere~\cite{Kirton2017}, signatures of the transition nonetheless survive.

\subsubsection{Linear stability analysis.}

To understand the role of the damping term $Q_1^\prime$ appearing in the equations of motion, we consider linear stability of the normal and superradiant states.

\paragraph{Normal state.}
For the normal state, $\langle S^x\rangle = \langle S^y \rangle=0, \langle S^z\rangle = - N/2$, we can easily see that fluctuations of $\langle S^z\rangle$ decouple from those of $\langle S^{x,y}\rangle$,
Denoting fluctuations of $\langle S^{x,y}\rangle$ by $x,y$, we find these obey the coupled equations:
\begin{equation}
    \partial_t 
    \begin{pmatrix} x \\ y \end{pmatrix}
    =
    \begin{pmatrix}
    0 & - \omega_0 \\
    \omega_0 + 4 Q_0^{\prime\prime} N & 4 Q_1^\prime N
    \end{pmatrix}
    \begin{pmatrix} x \\ y \end{pmatrix}.
\end{equation}
The eigenvalues of this matrix are:
\begin{equation}
    \lambda = 
    2 Q_1^\prime N 
    \pm i \sqrt{ \omega_0(\omega_0 + 4 Q_0^{\prime\prime} N) - (2 Q_1^\prime N)^2 }.
\end{equation}

We can use this to recover the behavior below and at threshold.  When $g$ is small, the square root is real, corresponding to oscillating modes, with damping caused by $Q_1^\prime$.  
Instability of the normal state occurs when an eigenvalue crosses zero.  This requires $\omega_0 + 4 Q_0^{\prime\prime} N = 0$.  Inserting the expression for $Q_0^{\prime\prime}$ we see this again matches the expected threshold, $g>g_c$ for superradiance of the Dicke model.  Using this definition of $g_c$ we find that away from this transition, we can write the eigenvalues as
\begin{equation}
    \lambda \simeq
       - \frac{ 4 g^2 N \kappa \omega \omega_0}{(\omega^2 + \kappa^2)^2}
       \pm i \omega_0 \sqrt{1 - \left(\frac{g}{g_c}\right)^2},
\end{equation}
which matches precisely the linear stability analysis of the full model as presented in Sec.~\ref{reviewDicke}.
For $g < g_c$, the eigenvalues have a small negative real part (given by the first term) and an imaginary part (given by the square root). Hence, the solution is stable. For $g > g_c$, one of the eigenvalues becomes positive. The solution $\langle S^z \rangle = -N/2$ is thus unstable. 

\paragraph{Superradiant state.}
One may similarly perform a linear stability analysis around the ordered states.
In this case we consider e.g. $\langle S^x\rangle = \langle S^x \rangle_{\text{ss}} + x$, where $\langle S^x \rangle_{\text{ss}}$ is the steady state value.  A similar replacement holds for $\langle S^z\rangle$, while  $\langle S^y \rangle = y$ as we may note that $\langle S^y\rangle_{\text{ss}} = 0$. This then gives:
\begin{equation}\label{eq:semiclassicallineraised}
    \begin{aligned}
    \partial_t x
    &= - \omega_0 y,
    \\
    \partial_t y
    &=
   \omega_0 x
    - 8 Q_0^{\prime\prime} \Big(
    \langle S^z \rangle_{\text{ss}} x 
    +
    \langle S^x \rangle_{\text{ss}} z
    \Big)
    - 8 Q_1^{\prime} \langle S^z \rangle_{\text{ss}} y,
    \\
    \partial_t z
    &=
     8 Q_0^{\prime\prime} \langle S^x \rangle_{\text{ss}} y.
    \end{aligned}
\end{equation}
The second equation can be simplified by noting that the steady state condition implies that $\omega_0 =8 Q_0^{\prime\prime} \langle S^z \rangle_{\text{ss}}$.  After this, one finds the equations for $y,z$ no longer depend on $x$, allowing one to directly read out the eigenvalues:
\begin{equation}
    \label{eq:sr_normal_modes}
    \lambda = -4 Q_1^\prime \langle S^z \rangle_{\text{ss}}
    \pm \sqrt{ (4 Q_1^\prime \langle S^z \rangle_{\text{ss}})^2 
    - (8  Q_0^{\prime\prime} \langle S^x \rangle_{\text{ss}})^2 }.
\end{equation}
We may note that since $Q_1^\prime \ll Q_0^{\prime\prime}$, the first term in square root can be neglected.  Inserting the expressions for the steady state values~(\ref{ssvalues}) we then find:
\begin{equation}
    \lambda \simeq - \frac{\kappa \omega_0^2}{\omega^2 + \kappa^2}
    \pm i \omega_0 \sqrt{\frac{g^4}{g_c^4}-1}.
\end{equation}
This once again matches the linear stability analysis of the full model reviewed in Sec.~\ref{reviewDicke}.

\section{Conclusion}
\label{sec:conclusion}

In summary, we have shown that it is possible to produce an atom-only description of the dissipative Dicke model which correctly describes the dissipation rates and the steady states of the system.  To recover such a theory it is necessary to use an unsecularized Redfield master equation, accounting for the variation of effective bath density of states at system frequencies $\pm \omega_0$.  Such a theory is not of Lindblad form.  Attempts to put it in Lindblad form, either by neglecting the small detuning $\omega_0$ or by secularization lead to a master equation that is qualitatively incorrect, i.e. it fails to produce the phase transition expected.  As a result, constructing a time-local equation of motion that captures the dissipative physics of the open Dicke model in an atom-only description appears to require such a non-Lindblad form. Although non-Lindblad master equations may lead to positivity violation, this violation of positivity is restricted to specific initial conditions, and does not typically occur for the steady state of the system.  Indeed, in other contexts~\cite{Eastham2016} it can be shown that violation of positivity of the density matrix only occurs as a transient behavior at early time starting from certain initial states;  this transient non-positivity does not cause problems for the later time evolution.

The ability to describe the dissipative dynamics with an atom-only theory is a crucial step to understand the quantum dynamics in more complicated situations. Specifically, for multimode cavities~\cite{Gop:Emergent,Gopalakrishnan2010a,Gopal:Frust,Gopalakrishnan:2012cf,Kollar2014,Kollar2016,Torggler2017,Vaidya2018,torggler2018quantum,guo2018sign,guo2018emergent}, it enables one to adiabatically eliminate the bosonic cavity modes (which massively increases the size of the Hilbert space), and provides a description of the key slow dynamics with only atomic variables.  Such a description can then form a basis for numerical methods, such as matrix-product-density-operator~\cite{Daley2004:MPDO,Verstraete2004:MPDO,Zwolak2004:MPDO}, corner-space-renormalization~\cite{Finazzi2015:Corner}, or cluster expansions~\cite{Jin2016:Cluster}, to be applied, allowing a full quantum description of the problem.  Moreover, as described above, the above quantum theory has a semiclassical limit that correctly captures the effects of dissipation.     In addition, the methods described here may be particularly useful when considering fermionic atoms in optical cavities~\cite{Keeling2014:Fermi,Piazza2014:Fermi,Chen2014:Fermi} and the possibility of cavity mediated superconductivity~\cite{Colella2018:SC}, as fermions do not admit the same semiclassical approaches as used for bosonic atoms. Future work will make use of the methods above applied to multimode problems to explore the evolution of entanglement and quantum correlations in this complex open quantum system.

All data underpinning this publication are openly available from the University of Strathclyde KnowledgeBase~\cite{data}.

\begin{acknowledgments}
J.~K.~acknowledges useful discussions with Monika Schleier-Smith and Benjamin Lev, and in particular acknowledges Monika Schleier-Smith for asking the question that prompted this work. 
F.~D.~and A.~D.~acknowledge support from the EPSRC Programme Grant DesOEQ (EP/P009565/1), and by the EOARD via AFOSR grant number FA9550-18-1-0064. 
J.~K.~acknowledges support from SU2P.
\end{acknowledgments}

\bibliography{bibliography}

\end{document}